\documentclass[twocolumn]{aastex62}
\usepackage{natbib}
\usepackage{epsfig}
\usepackage{graphicx}
\usepackage{subfigure}
\usepackage{float}
\usepackage{amsmath}
\usepackage{color}
\usepackage{amssymb}
\usepackage{amsfonts}
\usepackage{apjfonts}
\newcommand{\PMO}{Purple Mountain Observatory, Chinese Academy of Sciences, Nanjing 210034, China}
\newcommand{\GXU}{Guangxi Key Laboratory for Relativistic Astrophysics, Guangxi University, Nanning 530004, China}

\shortauthors{Wei}

\begin{document}

\title{Unbiased Cosmic Opacity Constraints from Standard Sirens and Candles}

\correspondingauthor{Jun-Jie Wei}
\email{jjwei@pmo.ac.cn}

\author{Jun-Jie Wei}
\affiliation{\PMO}
\affiliation{\GXU}

\begin{abstract}
The observation of Type Ia supernovae (SNe Ia) plays an essential role in probing the expansion history
of the universe. But the possible presence of cosmic opacity can degrade the quality of SNe Ia.
The gravitational-wave (GW) standard sirens, produced by the coalescence of double neutron stars and
black hole--neutron star binaries, provide an independent way to measure the distances of GW sources,
which are not affected by cosmic opacity. In this paper, we first propose that combining the GW observations
of third-generation GW detectors with SN Ia data in similar redshift ranges offers a novel and model-independent
method to constrain cosmic opacity. Through Monte Carlo simulations, we find that one can constrain
the cosmic opacity parameter $\kappa$ with an accuracy of $\sigma_{\kappa}\sim0.046$
by comparing the distances from 100 simulated GW events and 1048 current Pantheon SNe Ia. The uncertainty
of $\kappa$ can be further reduced to $\sim0.026$ if 800 GW events are considered. We also demonstrate
that combining 2000 simulated SNe Ia and 1000 simulated GW events could result in much severer
constraints on the transparent universe, for which $\kappa=0.0000\pm0.0044$. Compared to previous
opacity constraints involving distances from other cosmic probes, our method using GW standard sirens
and SN Ia standard candles at least achieves competitive results.
\end{abstract}


\section{Introduction}

In 1998, the accelerated expansion of the universe was first revealed by the unexpected dimming of
Type Ia supernovae (SNe Ia) \citep{1998AJ....116.1009R,1999ApJ...517..565P}. Soon after the discovery of cosmic acceleration,
a cosmological distribution of dust has been suggested as an alternative explanation for the observed
dimming of SNe Ia \citep{1999ApJ...512L..19A,1999ApJ...525..583A}. Indeed, SN observations are affected by
dust in the Milky Way, the intergalactic medium, intervening galaxies, and their host galaxies.
The extinction effects of SNe Ia caused by these dust in the Milky Way and their host galaxies
have been well modeled and they have no impact on the conclusion of cosmic acceleration. However,
cosmic opacity may also be due to other exotic mechanisms, in which extragalactic magnetic fields
turn photons into unobserved particles (e.g., light axions, gravitons, chameleons, Kaluza-Klein modes)
\citep{1995PhRvL..74..634C,2000PhRvD..62f3507D,2002PhRvL..88p1302C,2004PhRvL..93q1104K,2008PhRvD..77d3009B,2010JCAP...10..024A,2010ARNPS..60..405J}.
We have little knowledge about exotic mechanisms for cosmic opacity and their influence on SN observations.
Therefore, the question of whether cosmic opacity can be responsible for part of the dimming of
standard candles remains open. As more than 1000 SNe Ia have been detected \citep{2018ApJ...859..101S},
their cosmological constraint ability is now limited by systematic uncertainties rather than
by statistical errors. An important systematic uncertainty is the mapping of cosmic opacity.
In the era of precision cosmology, it is necessary to accurately quantify the transparency of the universe.

In the past, the cosmic distance duality (CDD) relation has been used to verify the presence of
opacity and systematic uncertainties in SN Ia data. The luminosity distance ($D_L$) and
the angular diameter distance ($D_A$) are related by the CDD relation
\citep{1933PMag...15..761E,2007GReGr..39.1047E}: $D_L=D_A(1+z)^{2}$.
This relation holds for all cosmological models described by Riemannian geometry, requiring that
photons travel along null geodesics and the number of photons is conserved \citep{2007GReGr..39.1047E}.
Many works have been performed to test the validity of the CDD relation (e.g., \citealt{2004PhRvD..69j1305B,
2004PhRvD..70h3533U,2010ApJ...722L.233H,2011A&A...528L..14H,2012A&A...538A.131H,2011PhRvL.106v1301K,
2011ApJ...729L..14L,2011JCAP...05..023N,2012MNRAS.420L..43G,2012ApJ...745...98M,2013PhRvD..87j3530E,
2013ApJ...777L..24Y,2016ApJ...822...74L,2016PDU....13..139L,2016JCAP...07..026R,2017arXiv171010929Y,
2018MNRAS.477.5064H,2018MNRAS.480.3117L,2018ApJ...861..124M,2018MNRAS.481.4855M,2018ApJ...866...31R}).
Meanwhile, assuming the deviation from the CDD relation is attributed to the non-conservation of
photon number, the opacity of the universe has been widely tested with astronomical observations
(e.g., \citealt{2009JCAP...06..012A,2010JCAP...10..024A,2009ApJ...696.1727M,2011ApJ...742L..26L,
2012JCAP...10..029C,2012JCAP...12..028N,2013JCAP...04..027H,2013PhRvD..87j3013L,2013PhLB..718.1166L,
2015PhRvD..92l3539L,2014PhRvD..89j3517H,2017ApJ...836..107H,2017GReGr..49..150J,2017ApJ...847...45W}).
In these works, some tests of cosmic opacity were carried out by adopting specific cosmological models
and others were given in a model-independent way. There are two general methods to obtain
model-independent constraints on cosmic opacity. The first is to confront the luminosity distances
inferred from SN Ia observations with the opacity-independent angular diameter distances derived from
baryon acoustic oscillations or galaxy clusters
\citep{2009ApJ...696.1727M,2012JCAP...10..029C,2012JCAP...12..028N,2013PhRvD..87j3013L}.
The other model-independent method was proposed by comparing the luminosity distances of SNe Ia with
the opacity-free luminosity distances obtained from the Hubble parameter $H(z)$ or the ages of old passive galaxies
\citep{2013JCAP...04..027H,2013PhLB..718.1166L,2015PhRvD..92l3539L,2017GReGr..49..150J,2017ApJ...847...45W}.

On the other hand, because the waveform signals of gravitational waves (GWs) from inspiralling
and merging compact binaries encode $D_L$ information \citep{1986Natur.323..310S},
one may construct the $D_L-z$ relation to probe cosmology if their electromagnetic (EM)
counterparts with known redshifts can be detected (see also~\citealt{2005ApJ...629...15H,2014PhRvX...4d1004M,2018PhRvD..97f4031Z}).
GWs are therefore deemed as standard sirens, analogous to SN Ia standard candles.
Recently, the detection of the GW event GW170817 coincident with its EM
counterparts from a binary neutron star (NS) merger provided a standard-siren measurement
of the Hubble constant $H_0$ \citep{2017Natur.551...85A}. In addition to measuring $H_0$,
other cosmological applications of future GW data have also been explored, such as
constraining the cosmological parameters and the nature of dark energy (e.g.,
~\citealt{2005ApJ...629...15H,2011PhRvD..83b3005Z,2012PhRvD..86d3011D,2017PhRvD..95d4024C,2017PhRvD..95d3502D,2018arXiv181201440D,2018ApJ...860L...7W}),
probing the CDD relation \citep{2017arXiv171010929Y}, testing the anisotropy of the universe \citep{2018PhRvD..97j3005C,2018EPJC...78..356L},
weighing the total neutrino mass \citep{2018PhLB..782...87W}, estimating the time variation of Newton's constant $G$ \citep{2018JCAP...10..052Z},
and determining the cosmic curvature in a model-independent way \citep{2018ApJ...868...29W}.

Unlike the distance calibrations of SNe Ia that are affected by cosmic opacity, distance measurements
from GW observations have the advantage of being insensitive to the non-conservation of photon number.
Therefore, GW standard sirens provide a novel way to determine the opacity-independent $D_L$ of SNe Ia
at the same redshifts. In this paper, we first propose that unbiased cosmic opacity tests can be performed
by combining SN Ia and GW data in similar redshift ranges. We make a detailed research on what level of
opacity constraints may be achieved using future GW observations from the third-generation GW detectors
such as the Einstein Telescope (ET).

The rest of the paper is organized as follows. In Section~\ref{sec:SNe}, we describe the opacity dependence
of SN standard candles. In Section~\ref{sec:GW}, we give an overview of using GWs as standard sirens
in the potential ET observations. Unbiased cosmic opacity constraints from standard sirens and candles are
discussed in Section~\ref{sec:result}. Finally, conclusions are drawn in Section~\ref{sec:summary}.
Throughout we use the geometric unit $G=c=1$.

\section{Opacity dependence of SNe Ia}
\label{sec:SNe}

As pointed out by \cite{2009JCAP...06..012A}, the distance moduli derived from SNe Ia would be
systematically influenced if there was a source of ``photon absorption'' affecting the universe
transparency. Any effect reducing the photon number would dim the SN luminosity and increase its $D_L$.
If $\tau(z)$ represents the opacity from a source at $z$ to an observer at $z=0$ due to
extinction, the received flux from the source would be decreased by a factor $e^{-\tau(z)}$.
Thus, the observed luminosity distance ($D_{L,{\rm obs}}$) is related to the true luminosity distance
($D_{L,{\rm true}}$) by
\begin{equation}
D_{L,{\rm obs}}=D_{L,{\rm true}}e^{\frac{\tau(z)}{2}}\;.
\label{eq:DLobs}
\end{equation}
The observed distance modulus is then given by
\begin{equation}
\mu_{\rm obs}(z)=\mu_{\rm true}(z)+2.5\left(\log_{10}e\right)\tau(z)\;.
\label{eq:dmobs}
\end{equation}

In order to use the full redshift range of the available data, we adopt the following simple
parametrization for a deviation from the CDD relation \citep{2009JCAP...06..012A}
\begin{equation}
D_{L,{\rm obs}}=D_A\left(1+z\right)^{2+\kappa}\;,
\label{eq:eq3}
\end{equation}
where the parameter $\kappa$ reflects the degree of departure from transparency.
Combining Equations~(\ref{eq:DLobs}) and (\ref{eq:eq3}) we obtain the exact form of
the opacity depth,
\begin{equation}
\tau(z)=2\kappa\ln\left(1+z\right)\;.
\label{eq:tau}
\end{equation}
To better understand the physical meaning of a constraint on $\kappa$,
\cite{2009JCAP...06..012A} noted that for small $\kappa$ and $z\leq1$ Equation~(\ref{eq:tau})
is equivalent to adopting an optical depth parametrization $\tau(z)=2\kappa z$ or
$\tau(z)=(1+z)^{\alpha}-1$ with the correspondence $\alpha=2\kappa$.
While this identification is based on a Taylor expansion, \cite{2009JCAP...06..012A}
proved that the expansion is good to better than 20\% for the entire $\kappa$ range
and the redshift range considered.

In this work, we consider the largest SN Ia sample called Pantheon, which consists of
1048 SNe Ia in the redshift range $0.01 < z < 2.3$ \citep{2018ApJ...859..101S}.
The observed distance moduli of SNe can be calculated from the SALT2 light-curve fit parameters using the formula
\begin{equation}
\mu_{\rm obs}^{\rm SN}=m_{B}-M_{B}+\alpha x_{1}-\beta C+\Delta_{M}+\Delta_{B}\;,
\label{eq:mu}
\end{equation}
where $m_B$ is the observed \emph{B}-band apparent magnitude, $M_{B}$ is the absolute $B$-band
magnitude, $x_{1}$ and $C$ are, respectively, the light-curve stretch factor and color parameter,
$\Delta_{M}$ denotes a distance correction based on the host galaxy mass, and $\Delta_{B}$
represents a distance correction from various biases predicted from simulations. $\alpha$ and $\beta$
are nuisance parameters that describe the luminosity--stretch and luminosity--color relations.

Generally, the two nuisance parameters $\alpha$ and $\beta$ are determined by fitting simultaneously
with cosmological parameters in a specific cosmological model. In this sense, the derived distances
of SNe Ia are cosmological-model-dependent. To avoid this problem, \cite{2017ApJ...836...56K}
introduced the BEAMS with Bias Corrections (BBC) method to calibrated the SNe. This method relies
heavily on \cite{2011ApJ...740...72M} but involves extensive simulations to correct the SALT2 fit parameters
$m_B$, $x_{1}$, and $C$. The BBC fit creates a bin-averaged Hubble diagram from SN Ia data, and then
the nuisance parameters $\alpha$ and $\beta$ are determined by fitting to an arbitrary cosmological model,
which is referred as the reference cosmology. The reference cosmological model is required to well
describe the local shape of the Hubble diagram within each redshift bin. As long as the number of bins
is large enough, the fitted parameters $\alpha$ and $\beta$ will converge to consistent values,
which are independent of the reference cosmology \citep{2011ApJ...740...72M}.

The distances of the Pantheon SNe were calibrated after using SALT2 light-curve fitter,
then applying the BBC method to determine the nuisance parameters, and adding the distance bias corrections
\citep{2018ApJ...859..101S}. The corrected apparent magnitudes $m_{\rm corr}=\mu_{\rm obs}^{\rm SN}+M_{B}$
of the Pantheon data are reported in \cite{2018ApJ...859..101S}. Therefore, to obtain
the observed distance modulus $\mu_{\rm obs}^{\rm SN}$, we just need to subtract $M_{B}$ from
$m_{\rm corr}$ and no longer need to do the stretch and color corrections.
Considering the effect of cosmic opacity on standard candles, the true distance modulus
can be written as
\begin{equation}
\mu_{\rm true}^{\rm SN}(z)=m_{\rm corr}-M_{B}-5\kappa \log_{10}\left(1+z\right)\;,
\label{eq:mu-ture}
\end{equation}
where we emphasize that $\kappa$ and $M_B$ are the only two free parameters.

\section{GW standard sirens}
\label{sec:GW}

From the observations of GW signals, caused by the coalescence of compact binaries, one can obtain
an absolute measure of $D_L$. If compact binaries are black hole (BH)--NS or NS--NS binaries,
the source redshifts may be available from EM counterparts that associated
with the GW events \citep{2010ApJ...725..496N,2010CQGra..27u5006S,2011PhRvD..83b3005Z,2017PhRvD..95d4024C}.
Therefore, this offers a model-independent way to establish the $D_{L}$--$z$ relation (or the Hubble diagram)
over a wide redshift range. The ET, with the designed high-sensitivity (10 times more sensitive
in amplitude than current advanced laser interferometric detectors) and wide frequency range ($1-10^{4}$ Hz),
would be able to see NS--NS merger GW events up to redshifts of $z\sim2$ and BH--NS events up to $z>2$
\citep{2010CQGra..27s4002P}. In this section, we briefly summarize the method to simulate the GW data from the ET.

The first step for generating GW standard sirens is to simulate the redshift distribution of
the sources. Following \cite{2011PhRvD..83b3005Z} and \cite{2017PhRvD..95d4024C}, we expect
the source redshifts can be measured by identifying EM counterparts from the coalescence of
double NSs and BH--NS binaries. The redshift distribution of the observable sources
takes the form \citep{2011PhRvD..83b3005Z}
\begin{equation}
P(z)\propto \frac{4\pi D_C^2(z)R(z)}{H(z)(1+z)},
\label{equa:pz}
\end{equation}
where $D_C(z)$ denotes the comoving distance and $R(z)$ is the merger rate of binary
systems (BH--NS or NS--NS) with the expression \citep{2001MNRAS.324..797S,2009PhRvD..80j4009C}
\begin{equation}
R(z)=\begin{cases}
1+2z, & z\leq 1 \\
\frac{3}{4}(5-z), & 1<z<5 \\
0, & z\geq 5.
\end{cases}
\label{equa:rz}
\end{equation}
We simulate the source redshift $z$ according to this redshift distribution.
Note that although the Pantheon sample covers a wide redshift range of $0.01<z<2.3$,
there is only one SN located at $z>2$ \citep{2018ApJ...859..101S}.
To be consistent with the redshift range of the Pantheon SNe, we consider the potential
observations of GW events in $0< z < 2.0$. With the mock $z$, the fiducial luminosity distance
$D_{L}^{\rm fid}$ can be calculated in the fiducial flat $\Lambda$CDM model
\begin{equation}
D_{L}(z)=\frac{1+z}{H_0}\int_{0}^{z}\frac{{\rm d}z}{\sqrt{\Omega_{m} (1+z)^{3}+1-\Omega_{m}}}\;.
\end{equation}
Here we adopt the following cosmological parameters:
$H_{0}=70.0$ km $\rm s^{-1}$ $\rm Mpc^{-1}$ and $\Omega_{m}=0.298$
\citep{2018ApJ...859..101S}.

The next step is to get the total error $\sigma_{D_L}$ in the luminosity distance of the GW source.
In order to calculate $\sigma_{D_L}$, one needs to generate the waveform of GWs.
The detector response to a GW signal is a linear combination of two wave polarizations,
$h(t)=F_+h_+(t)+F_\times h_\times(t)$. The detector's antenna pattern functions $F_{+}$ and $F_\times$
depend on the source's position ($\theta_s, \phi_s$) and the polarization angle $\psi_s$.
The restricted post-Newtonian approximation waveforms $h_\times$ and $h_+$ for the non-spinning
compact binaries depend on the symmetric mass ratio $\eta=m_1 m_2/(m_1+m_2)^2$, the chirp mass
$\mathcal{M}_{c}=(m_1+m_2)\eta^{3/5}$ ($m_1$ and $m_2$ are component masses of a coalescing binary),
the inclination angle $\iota$ between the binary's orbital and the line-of-sight, the $D_L$,
the epoch of the merger $t_0$, and the merging phase $\psi_0$ \citep{2009LRR....12....2S}.
So, for a given binary, the response of the detector depends on $(\mathcal{M}_{c},\;\eta,\;
t_0,\;\psi_0,\;\theta_s,\;\phi_s,\;\psi_s,\;\iota,\;D_L)$. Using the Fisher information matrix
and marginalizing over the other parameters, we can estimate the instrumental error $\sigma_{D_L}^{\rm inst}$
on the measurement of $D_L$. In addition to $\sigma_{D_L}^{\rm inst}$, we also consider an error
$\sigma_{D_L}^{\rm lens}/D_L=0.05z$ due to the weak lensing effect. Thus, the total
uncertainty is $\sigma_{D_L}=\left[(\sigma_{D_L}^{\rm inst})^{2}+(\sigma_{D_L}^{\rm lens})^{2}\right]^{1/2}$.
Readers may refer to \cite{2017PhRvD..95d4024C} for detailed information about the production
of $\sigma_{D_L}$ (see also \citealt{2011PhRvD..83b3005Z,2018PhLB..782...87W,2018ApJ...860L...7W,2018ApJ...868...29W}).
Note that the signal is identified as a GW detection only when the evaluated signal-to-noise ratio (S/N)
is larger than $8.0$. For every confirmed detection (i.e., S/N$>8.0$), the fiducial luminosity distance
$D_{L}^{\rm fid}$ is converted to the fiducial distance modulus by
\begin{equation}
\mu^{\rm fid}=5 \log_{10}\left(\frac{D_{L}^{\rm fid}}{\rm Mpc}\right)+25\;,
\end{equation}
and the error of $\mu^{\rm fid}$ is propagated from that of $D_{L}^{\rm fid}$ by
\begin{equation}
\sigma_{\mu^{\rm GW}}=\frac{5}{\ln10}\frac{\sigma_{D_L}}{D_{L}^{\rm fid}}\;.
\end{equation}
We then add the deviation $\sigma_{\mu^{\rm GW}}$ to the fiducial value of $\mu^{\rm fid}$.
That is, we sample the $\mu^{\rm GW}$ measurement according to the Gaussian distribution
$\mu^{\rm GW}=\mathcal{N}(\mu^{\rm fid},\;\sigma_{\mu^{\rm GW}})$.

\begin{figure}
\vskip-0.2in
\centerline{\includegraphics[angle=0,width=1.15\hsize]{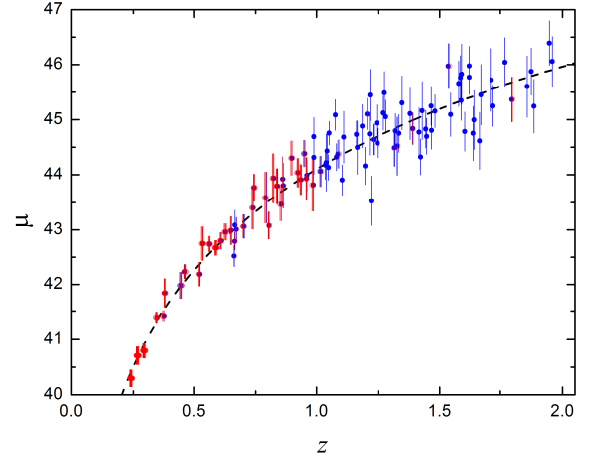}}
\vskip-0.2in
\caption{Example catalog of 100 simulated GW events (blue dots) with redshifts $z$, distance moduli $\mu$,
and the errors in the distance moduli $\sigma_{\mu}$. The dashed line is the fiducial flat $\Lambda$CDM model.
Red circles represent 213 Pantheon SNe Ia whose distance moduli are determined by the nearest GW data,
other 835 SNe Ia that have redshift differences $\Delta z\geq0.005$ with respect to their corresponding nearest GW data are discarded.}
\label{f1}
\end{figure}

Using the method described above, one can generate a catalog of the simulated GW events with
$z$, $\mu^{\rm GW}$, and $\sigma_{\mu^{\rm GW}}$. As argued in \cite{2017PhRvD..95d4024C},
the ET is expected to detect $\mathcal{O}(10^2)$ GW sources with EM counterparts per year.
Thus, we first simulate a population of 100 such events. An example of 100 simulated GW data
(blue dots) from the fiducial model is presented in Figure~\ref{f1}.

\section{Unbiased constraints on cosmic opacity}
\label{sec:result}
Future detectable GW sources are expected to distribute in nearly the same redshift range as the SN Ia data,
and the opacity-free $\mu^{\rm GW}$ of GW events can be provided by the GW observations alone.
By confronting distance muduli $\mu^{\rm GW}(z)$ from the simulated GW events with distance muduli
$\mu_{\rm true}^{\rm SN}(z)$ in Equation~(\ref{eq:mu-ture}) that depend on $\kappa$ and $M_B$ from
observations of SNe Ia at the same redshifts, we can obtain a model-independent constraint on cosmic opacity.
However, in reality, it is difficult to have both $\mu^{\rm GW}$ and $\mu_{\rm true}^{\rm SN}$ at
exactly the same redshift. So, as \cite{2010ApJ...722L.233H} and \cite{2011ApJ...729L..14L} did in their
treatments, we find the nearest redshift to GW data from SNe Ia and use the criteria
($\Delta z=|z_{\rm GW}-z_{\rm SN}|<0.005$) to ensure the redshift differences of the nearest SNe Ia
to GW data are not too large. For the example of 100 simulated GW data shown in Figure~\ref{f1},
we find that there are 213 Pantheon SNe Ia (red circles) that satisfy the redshift selection criteria.
Other 835 SNe Ia that have redshift differences $\Delta z\geq0.005$ are discarded.

\begin{figure}
\vskip-0.2in
\centerline{\includegraphics[keepaspectratio,clip,width=0.55\textwidth]{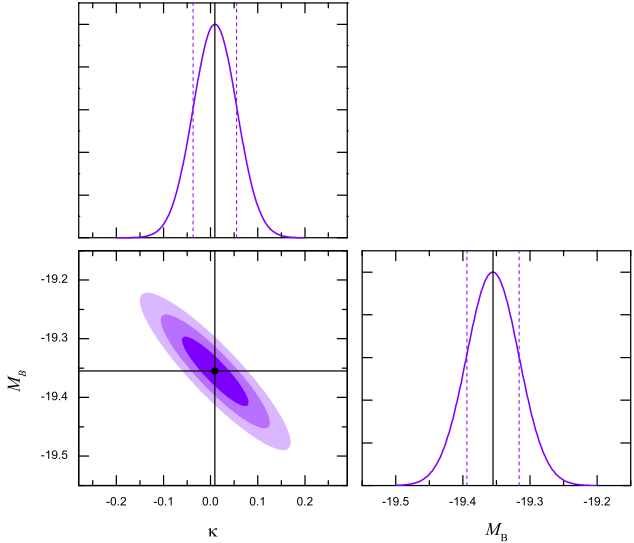}}
\vskip-0.2in
\caption{1-D marginalized probability distributions and $1-3\sigma$ constraint contours for
cosmic opacity $\kappa$ and SN Ia absolute magnitude $M_B$, using 100 simulated GW events
and observations of Pantheon SNe Ia. The vertical solid lines represent the best-fit values,
and the vertical dashed lines enclose the $1\sigma$ credible region.}
\label{f2}
\end{figure}

We now give the $\chi^{2}$ statistic for constraining cosmic opacity parameterized by $\kappa$,
\begin{equation}
\chi^2(\kappa,\;M_B)=\sum_{i}\frac{\left[\mu_{\rm true}^{\rm SN}(z_{i};\;\kappa,\;M_B)-\mu^{\rm GW}(z_{i})\right]^{2}}
{\sigma_{\mu^{\rm SN},i}^{2}+\sigma_{\mu^{\rm GW},i}^{2}}\;,
\end{equation}
where $\sigma_{\mu^{\rm SN}}$ is the observational error of the SN distance modulus. Here only
the statistical uncertainties are considered since only part of Pantheon SNe Ia are
selected to match the simulated GW data. To make sure the final constraints are unbiased,
we repeat this process 1000 times for each GW data set using different noise seeds.
Figure~\ref{f2} displays the constraint results on $\kappa$ and $M_B$.
We find that, from 100 simulated GW events and observations of Pantheon SNe Ia,
the unbiased constraint on cosmic opacity is $\kappa=0.009\pm0.046$ ($1\sigma$).\footnote{
After this work appeared on arXiv, we found a similar work \citep{2019arXiv190201702Q},
which has also independently investigated opacity constraints from GWs and SNe Ia.}

\begin{figure}
\vskip-0.1in
\centerline{\includegraphics[keepaspectratio,clip,width=0.55\textwidth]{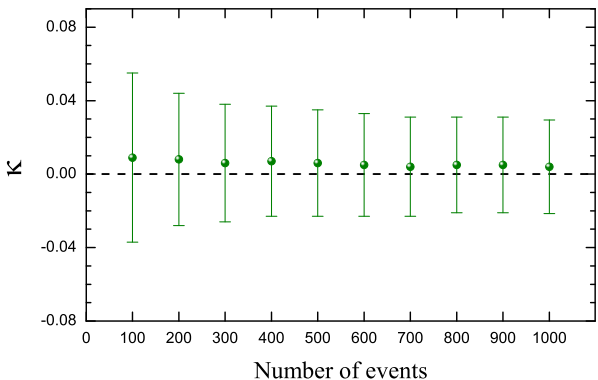}}
\vskip-0.1in
\caption{Best-fit cosmic opacity $\kappa$ and $1\sigma$ confidence level as a function of the number of GW events.
The dashed line corresponds to a transparent universe.}
\label{f3}
\end{figure}

Note that the number of observable GW events is quite uncertain. To test how the uncertainty
of $\kappa$ depends on the number of simulated GW events ($N_{\rm GW}$),
in Figure~\ref{f3} and Table~\ref{table1} we show the best-fit $\kappa$ and $1\sigma$ confidence level
as a function of $N_{\rm GW}$. One can see from Figure~\ref{f3} and Table~\ref{table1}
that the uncertainty of $\kappa$ is gradually reduced with the increasing of the number of GW events,
finally turns to a relatively stable value (i.e., $\sigma_{\kappa}\simeq0.026$).
The constraint results are nearly the same for the cases of $N_{\rm GW}\geq800$, which is understandable.
We only use the data of GWs and SNe Ia that satisfying the criteria ($\Delta z=|z_{\rm GW}-z_{\rm SN}|<0.005$)
to constrain $\kappa$. With the fixed Pantheon SN sample ($N_{\rm SN}=1048$), the number of GW/SN pairs satisfying
the redshift selection criteria would begin to stabilize and the resulting constraints
would be nearly the same, when the number of GW events is large enough.

\begin{table}
\centering \caption{Summary of Unbiased Cosmic Opacity Constraints from
$N_{\rm GW}$ Simulated GW Events and Observations of Pantheon SNe Ia}
\begin{tabular}{cc|cc}
\hline
\hline
 $N_{\rm GW}$ &  $\kappa$  &  $N_{\rm GW}$  &  $\kappa$ \\
\hline
100   &   $0.009\pm0.046$   &   600   &   $0.005\pm0.028$   \\
200   &   $0.008\pm0.036$   &   700   &   $0.004\pm0.027$   \\
300   &   $0.006\pm0.032$   &   800   &   $0.005\pm0.026$   \\
400   &   $0.007\pm0.030$   &   900   &   $0.005\pm0.026$   \\
500   &   $0.006\pm0.029$   &   1000  &   $0.004\pm0.026$   \\
\hline
\end{tabular}
\label{table1}
\end{table}

\begin{figure}
\vskip-0.5in
\centerline{\includegraphics[keepaspectratio,clip,width=0.5\textwidth]{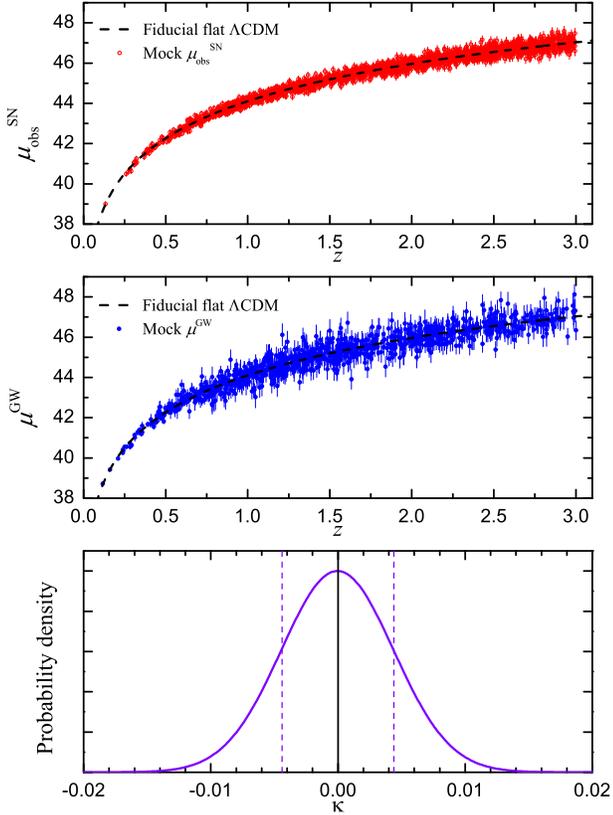}}
\vskip-0.7in
\caption{An example of the simulations for the case of 2000 simulated SNe Ia and 1000 simulated GW events.
Top panel shows the Hubble diagram of 2000 simulated SNe Ia with the fiducial flat $\Lambda$CDM model (dashed line).
Middle panel shows the Hubble diagram of 1000 simulated GW events with the fiducial flat $\Lambda$CDM model (dashed line).
Bottom panel shows the final constraint on cosmic opacity $\kappa$ from these data.}
\label{f4}
\end{figure}

By the time we have ET results, much more SN Ia data with wider redshift range
may be detected by future SN surveys. It is expected that more than 2000 SNe Ia
can be detected in the era of the \emph{Wide Field Infrared Survey Telescope}
(\emph{WFIRST}) \citep{2012arXiv1208.4012G}. To better represent how effective our
method might be with more SN Ia measurements, we also perform Monte Carlo simulations
to create the mock $\mu^{\rm SN}_{\rm obs}-z$ data sets. We assume that there are
2000 SNe Ia by the time that 1000 GW events are detected. The redshift distribution
of SNe Ia is adopted as
\begin{equation}
P_{\rm SN}(z)\propto \frac{4\pi D_C^2(z)R_{\rm SN}(z)}{H(z)(1+z)},
\label{equa:pSN}
\end{equation}
where $R_{\rm SN}(z)$ is the volumetric rate of SNe Ia, which is given by \citep{2018ApJ...867...23H}
\begin{equation}
R_{\rm SN}(z)=\begin{cases}
2.5\times\left(1+z\right)^{1.5}\left(10^{-5}\;{\rm Mpc^{-3}}\;{\rm yr^{-1}}\right), & {\rm for}\;z\leq 1 \\
5.0\times\left(1+z\right)^{0.5}\left(10^{-5}\;{\rm Mpc^{-3}}\;{\rm yr^{-1}}\right), & {\rm for}\;1<z<3.
\end{cases}
\label{equa:rSN}
\end{equation}
As the expected detection rate for $z>3$ SNe is low, we do not attempt to simulate
SNe at those redshifts. Following \cite{2018ApJ...867...23H}, the total distance uncertainty $\sigma_{\mu^{\rm SN}}$
of each mock SN is calculated by the sum of the systematic uncertainty $\sigma_{\rm sys}$ and
the statistical uncertainty $\sigma_{\rm stat}$, i.e., $\sigma_{\mu^{\rm SN}}^{2}=\sigma_{\rm sys}^{2}+\sigma_{\rm stat}^{2}$.
The systematic uncertainty is assumed to increase with redshift, $\sigma_{\rm sys}=\frac{0.01(1+z)}{1.8}$ (mag).
The statistical uncertainty is $\sigma_{\rm stat}^{2}=\sigma_{\rm meas}^{2}+\sigma_{\rm int}^{2}+\sigma_{\rm lens}^{2}$,
where $\sigma_{\rm meas}=0.08$ mag includes both statistical measurement uncertainties and
statistical model uncertainties, $\sigma_{\rm int}=0.09$ mag denotes the intrinsic scatter
in the corrected SN Ia distances, and $\sigma_{\rm lens}=0.07\times z$ mag represents the lensing
uncertainty. The route of GW simulation is the same as described earlier in Section~\ref{sec:GW},
but now we consider the potential observations of GW events in $0 < z < 3.0$. Figure~\ref{f4}
gives an example of the simulations for the case of 2000 simulated SNe Ia and 1000 simulated GW events.
From top to bottom, the three panels show the Hubble diagram of 2000 simulated SNe Ia with
the fiducial flat $\Lambda$CDM model (dashed line), the Hubble diagram of 1000 simulated GW events
with the fiducial flat $\Lambda$CDM model (dashed line), and the final constraint on $\kappa$, respectively.
In this case, the final derived $\kappa$ is $\kappa=0.0000\pm0.0044$ ($1\sigma$).
Compared with the constraint obtained from 1048 Pantheon SNe Ia and 800 simulated GW events
($\kappa=0.005\pm0.026$), the uncertainty of the determined $\kappa$
in this case can be further improved by a factor of $\sim6.0$.

In our above simulations, the fiducial model is chosen to be flat $\Lambda$CDM.
To investigate a possible degeneracy of the results with the adopted fiducial model, we also
consider two separate cosmological models: $w$CDM and nonflat $\Lambda$CDM. We take
the best-fit cosmological parameters from the Pantheon SN sample \citep{2018ApJ...859..101S}
as the fiducial models ($w$CDM: $\Omega_{m}=0.316$ and $w=-1.09$; nonflat $\Lambda$CDM:
$\Omega_{m}=0.319$ and $\Omega_{\Lambda}=0.733$) to generate 2000 simulated SNe Ia and
1000 simulated GW events in $0 < z < 3.0$. The final opacity constraints are $\kappa=-0.0003\pm0.0044$
and $\kappa=-0.0004\pm0.0044$ for the fiducial $w$CDM and nonflat $\Lambda$CDM
models, respectively. By comparing these constraints with that obtained from the fiducial
flat $\Lambda$CDM model ($\kappa=0.0000\pm0.0044$), we can conclude that the opacity constraints
are independent of the adopted fiducial model.

\section{Summary and discussion}
\label{sec:summary}
Cosmic opacity may be due to absorption or scattering caused by dust in the universe,
or may result from other exotic mechanisms where extragalactic magnetic fields turn
photons into unobserved particles (e.g., light axions, gravitons, chameleons, Kaluza-Klein modes).
The presence of cosmic opacity can lead to a significant deviation from photon number conservation,
thus making the observed SNe Ia dimmer than what expected and affecting the reliable
reconstruction of the cosmic expansion history. It is therefore crucial to quantitatively
study the effect of cosmic opacity on SN standard candles.

As the luminosity distances of GWs can be directly obtained from their waveform signals
rather than from luminosities, they are independent of the non-conservation of photon number
and thus are not affected by cosmic opacity. In this work, we first propose that combining
the GW observations with SN Ia data in similar redshift ranges provides a novel
way to constrain cosmic opacity. Unbiased cosmic opacity constraints
are performed by comparing two kinds of distance moduli obtained from the recent Pantheon
compilation of SN Ia data and future GW observations of the ET. Our simulations show that,
from 1048 SN Ia measurements and 100 simulated GW events, the cosmic opacity parameter $\kappa$
is expected to be constrained with an accuracy of $\sigma_{\kappa}\sim0.046$. If 800 GW events
are observed, the uncertainty of $\kappa$ can be further reduced to $\sim0.026$.
We also demonstrate that with 2000 simulated SNe Ia and 1000 simulated GW events,
one can expect the transparent universe to be estimated at the precision of $\kappa=0.0000\pm0.0044$.

Previously, \cite{2009JCAP...06..012A,2010JCAP...10..024A} obtained $\kappa=-0.01^{+0.08}_{-0.09}$
and $\kappa=-0.04^{+0.08}_{-0.07}$ by analyzing SN Ia and $H(z)$ data in the flat $\Lambda$CDM model. From the joint analyses
involving gamma-ray bursts and $H(z)$ measurements, \cite{2014PhRvD..89j3517H}
obtained $\kappa=0.06^{+0.18}_{-0.18}$ and $\kappa=0.057^{+0.21}_{-0.21}$ in the
flat $\Lambda$CDM and XCDM frameworks, respectively. \cite{2015PhRvD..92l3539L}
obtained a model-independent constraint on cosmic opacity ($\kappa=0.07^{+0.11}_{-0.12}$)
by comparing the distances from SN Ia and $H(z)$ observations. \cite{2017GReGr..49..150J}
tested the conservation of photon number with distances from SNe Ia and those inferred
from ages of 32 old objects, yielding $\kappa=-0.18^{+0.25}_{-0.24}$.
By comparing our results with previous opacity constraints involving distances from different observations,
we prove that our method using GW standard sirens and SN Ia standard candles will be competitive.
Most importantly, our method offers a new model-independent way to constrain cosmic opacity.

\acknowledgments
We are grateful to the anonymous referee for helpful comments.
This work is partially supported by the National Natural Science Foundation of China
(grant Nos. U1831122, 11603076, 11673068, and 11725314), the Youth Innovation Promotion
Association (2017366), the Key Research Program of Frontier Sciences (grant No. QYZDB-SSW-SYS005),
the Strategic Priority Research Program ``Multi-waveband gravitational wave Universe''
(grant No. XDB23000000) of Chinese Academy of Sciences, and the ``333 Project''
and the Natural Science Foundation (grant No. BK20161096) of Jiangsu Province.


\begin{thebibliography}{}
\expandafter\ifx\csname natexlab\endcsname\relax\def\natexlab#1{#1}\fi

\bibitem[{{Abbott} {et~al.}(2017){Abbott}, {Abbott}, {Abbott}, {Acernese},
  {Ackley}, {Adams}, {Adams}, {Addesso}, {Adhikari}, {Adya}, \&
  et~al.}]{2017Natur.551...85A}
{Abbott}, B.~P., {Abbott}, R., {Abbott}, T.~D., {et~al.} 2017, \nat, 551, 85

\bibitem[{{Aguirre}(1999{\natexlab{a}})}]{1999ApJ...512L..19A}
{Aguirre}, A.~N. 1999{\natexlab{a}}, \apjl, 512, L19

\bibitem[{{Aguirre}(1999{\natexlab{b}})}]{1999ApJ...525..583A}
{Aguirre}, A. 1999{\natexlab{b}}, \apj, 525, 583

\bibitem[{{Avgoustidis} {et~al.}(2010){Avgoustidis}, {Burrage}, {Redondo},
  {Verde}, \& {Jimenez}}]{2010JCAP...10..024A}
{Avgoustidis}, A., {Burrage}, C., {Redondo}, J., {Verde}, L., \& {Jimenez}, R.
  2010, \jcap, 10, 024

\bibitem[{{Avgoustidis} {et~al.}(2009){Avgoustidis}, {Verde}, \&
  {Jimenez}}]{2009JCAP...06..012A}
{Avgoustidis}, A., {Verde}, L., \& {Jimenez}, R. 2009, \jcap, 6, 012

\bibitem[{{Bassett} \& {Kunz}(2004)}]{2004PhRvD..69j1305B}
{Bassett}, B.~A., \& {Kunz}, M. 2004, \prd, 69, 101305

\bibitem[{{Burrage}(2008)}]{2008PhRvD..77d3009B}
{Burrage}, C. 2008, \prd, 77, 043009

\bibitem[{{Cai} {et~al.}(2018){Cai}, {Liu}, {Liu}, {Wang}, \&
  {Yang}}]{2018PhRvD..97j3005C}
{Cai}, R.-G., {Liu}, T.-B., {Liu}, X.-W., {Wang}, S.-J., \& {Yang}, T. 2018,
  \prd, 97, 103005

\bibitem[{{Cai} \& {Yang}(2017)}]{2017PhRvD..95d4024C}
{Cai}, R.-G., \& {Yang}, T. 2017, \prd, 95, 044024

\bibitem[{{Chen} {et~al.}(2012){Chen}, {Wu}, {Yu}, \&
  {Li}}]{2012JCAP...10..029C}
{Chen}, J., {Wu}, P., {Yu}, H., \& {Li}, Z. 2012, \jcap, 10, 029

\bibitem[{{Chen}(1995)}]{1995PhRvL..74..634C}
{Chen}, P. 1995, Physical Review Letters, 74, 634

\bibitem[{{Cs{\'a}ki} {et~al.}(2002){Cs{\'a}ki}, {Kaloper}, \&
  {Terning}}]{2002PhRvL..88p1302C}
{Cs{\'a}ki}, C., {Kaloper}, N., \& {Terning}, J. 2002, Physical Review Letters,
  88, 161302

\bibitem[{{Cutler} \& {Holz}(2009)}]{2009PhRvD..80j4009C}
{Cutler}, C., \& {Holz}, D.~E. 2009, \prd, 80, 104009

\bibitem[{{Deffayet} \& {Uzan}(2000)}]{2000PhRvD..62f3507D}
{Deffayet}, C., \& {Uzan}, J.-P. 2000, \prd, 62, 063507

\bibitem[{{Del Pozzo}(2012)}]{2012PhRvD..86d3011D}
{Del Pozzo}, W. 2012, \prd, 86, 043011

\bibitem[{{Del Pozzo} {et~al.}(2017){Del Pozzo}, {Li}, \&
  {Messenger}}]{2017PhRvD..95d3502D}
{Del Pozzo}, W., {Li}, T.~G.~F., \& {Messenger}, C. 2017, \prd, 95, 043502

\bibitem[{{Du} {et~al.}(2018){Du}, {Yang}, {Xu}, {Pan}, \&
  {Mota}}]{2018arXiv181201440D}
{Du}, M., {Yang}, W., {Xu}, L., {Pan}, S., \& {Mota}, D.~F. 2018, arXiv
  e-prints, arXiv:1812.01440

\bibitem[{{Ellis}(2007)}]{2007GReGr..39.1047E}
{Ellis}, G.~F.~R. 2007, General Relativity and Gravitation, 39, 1047

\bibitem[{{Ellis} {et~al.}(2013){Ellis}, {Poltis}, {Uzan}, \&
  {Weltman}}]{2013PhRvD..87j3530E}
{Ellis}, G.~F.~R., {Poltis}, R., {Uzan}, J.-P., \& {Weltman}, A. 2013, \prd,
  87, 103530

\bibitem[{{Etherington}(1933)}]{1933PMag...15..761E}
{Etherington}, I.~M.~H. 1933, Philosophical Magazine, 15

\bibitem[{{Gon{\c c}alves} {et~al.}(2012){Gon{\c c}alves}, {Holanda}, \&
  {Alcaniz}}]{2012MNRAS.420L..43G}
{Gon{\c c}alves}, R.~S., {Holanda}, R.~F.~L., \& {Alcaniz}, J.~S. 2012, \mnras,
  420, L43

\bibitem[{{Green} {et~al.}(2012){Green}, {Schechter}, {Baltay}, {Bean},
  {Bennett}, {Brown}, {Conselice}, {Donahue}, {Fan}, {Gaudi}, {Hirata},
  {Kalirai}, {Lauer}, {Nichol}, {Padmanabhan}, {Perlmutter}, {Rauscher},
  {Rhodes}, {Roellig}, {Stern}, {Sumi}, {Tanner}, {Wang}, {Weinberg}, {Wright},
  {Gehrels}, {Sambruna}, {Traub}, {Anderson}, {Cook}, {Garnavich},
  {Hillenbrand}, {Ivezic}, {Kerins}, {Lunine}, {McDonald}, {Penny}, {Phillips},
  {Rieke}, {Riess}, {van der Marel}, {Barry}, {Cheng}, {Content}, {Cutri},
  {Goullioud}, {Grady}, {Helou}, {Jackson}, {Kruk}, {Melton}, {Peddie},
  {Rioux}, \& {Seiffert}}]{2012arXiv1208.4012G}
{Green}, J., {Schechter}, P., {Baltay}, C., {et~al.} 2012, arXiv e-prints,
  arXiv:1208.4012

\bibitem[{{Holanda} \& {Busti}(2014)}]{2014PhRvD..89j3517H}
{Holanda}, R.~F.~L., \& {Busti}, V.~C. 2014, \prd, 89, 103517

\bibitem[{{Holanda} {et~al.}(2013){Holanda}, {Carvalho}, \&
  {Alcaniz}}]{2013JCAP...04..027H}
{Holanda}, R.~F.~L., {Carvalho}, J.~C., \& {Alcaniz}, J.~S. 2013, \jcap, 4, 027

\bibitem[{{Holanda} {et~al.}(2010){Holanda}, {Lima}, \&
  {Ribeiro}}]{2010ApJ...722L.233H}
{Holanda}, R.~F.~L., {Lima}, J.~A.~S., \& {Ribeiro}, M.~B. 2010, \apjl, 722,
  L233

\bibitem[{{Holanda} {et~al.}(2011){Holanda}, {Lima}, \&
  {Ribeiro}}]{2011A&A...528L..14H}
{Holanda}, R.~F.~L., {Lima}, J.~A.~S., \& {Ribeiro}, M.~B. 2011, \aap, 528, L14

\bibitem[{{Holanda} {et~al.}(2012){Holanda}, {Lima}, \&
  {Ribeiro}}]{2012A&A...538A.131H}
{Holanda}, R.~F.~L., {Lima}, J.~A.~S., \& {Ribeiro}, M.~B. 2012, \aap, 538, A131

\bibitem[{{Holz} \& {Hughes}(2005)}]{2005ApJ...629...15H}
{Holz}, D.~E., \& {Hughes}, S.~A. 2005, \apj, 629, 15

\bibitem[{{Hounsell} {et~al.}(2018){Hounsell}, {Scolnic}, {Foley}, {Kessler},
  {Miranda}, {Avelino}, {Bohlin}, {Filippenko}, {Frieman}, {Jha}, {Kelly},
  {Kirshner}, {Mandel}, {Rest}, {Riess}, {Rodney}, \&
  {Strolger}}]{2018ApJ...867...23H}
{Hounsell}, R., {Scolnic}, D., {Foley}, R.~J., {et~al.} 2018, \apj, 867, 23

\bibitem[{{Hu} \& {Wang}(2018)}]{2018MNRAS.477.5064H}
{Hu}, J., \& {Wang}, F.~Y. 2018, \mnras, 477, 5064

\bibitem[{{Hu} {et~al.}(2017){Hu}, {Yu}, \& {Wang}}]{2017ApJ...836..107H}
{Hu}, J., {Yu}, H., \& {Wang}, F.~Y. 2017, \apj, 836, 107

\bibitem[{{Jaeckel} \& {Ringwald}(2010)}]{2010ARNPS..60..405J}
{Jaeckel}, J., \& {Ringwald}, A. 2010, Annual Review of Nuclear and Particle
  Science, 60, 405

\bibitem[{{Jesus} {et~al.}(2017){Jesus}, {Holanda}, \&
  {Dantas}}]{2017GReGr..49..150J}
{Jesus}, J.~F., {Holanda}, R.~F.~L., \& {Dantas}, M.~A. 2017, General
  Relativity and Gravitation, 49, 150

\bibitem[{{Kessler} \& {Scolnic}(2017)}]{2017ApJ...836...56K}
{Kessler}, R., \& {Scolnic}, D. 2017, \apj, 836, 56

\bibitem[{{Khedekar} \& {Chakraborti}(2011)}]{2011PhRvL.106v1301K}
{Khedekar}, S., \& {Chakraborti}, S. 2011, Physical Review Letters, 106, 221301

\bibitem[{{Khoury} \& {Weltman}(2004)}]{2004PhRvL..93q1104K}
{Khoury}, J., \& {Weltman}, A. 2004, Physical Review Letters, 93, 171104

\bibitem[{{Li} {et~al.}(2011){Li}, {Wu}, \& {Yu}}]{2011ApJ...729L..14L}
{Li}, Z., {Wu}, P., \& {Yu}, H. 2011, \apjl, 729, L14

\bibitem[{{Li} {et~al.}(2013){Li}, {Wu}, {Yu}, \& {Zhu}}]{2013PhRvD..87j3013L}
{Li}, Z., {Wu}, P., {Yu}, H., \& {Zhu}, Z.-H. 2013, \prd, 87, 103013

\bibitem[{{Liao} {et~al.}(2015){Liao}, {Avgoustidis}, \&
  {Li}}]{2015PhRvD..92l3539L}
{Liao}, K., {Avgoustidis}, A., \& {Li}, Z. 2015, \prd, 92, 123539

\bibitem[{{Liao} {et~al.}(2016){Liao}, {Li}, {Cao}, {Biesiada}, {Zheng}, \&
  {Zhu}}]{2016ApJ...822...74L}
{Liao}, K., {Li}, Z., {Cao}, S., {et~al.} 2016, \apj, 822, 74

\bibitem[{{Liao} {et~al.}(2013){Liao}, {Li}, {Ming}, \&
  {Zhu}}]{2013PhLB..718.1166L}
{Liao}, K., {Li}, Z., {Ming}, J., \& {Zhu}, Z.-H. 2013, Physics Letters B, 718,
  1166

\bibitem[{{Lima} {et~al.}(2011){Lima}, {Cunha}, \&
  {Zanchin}}]{2011ApJ...742L..26L}
{Lima}, J.~A.~S., {Cunha}, J.~V., \& {Zanchin}, V.~T. 2011, \apjl, 742, L26

\bibitem[{{Lin} {et~al.}(2018{\natexlab{a}}){Lin}, {Li}, \&
  {Li}}]{2018MNRAS.480.3117L}
{Lin}, H.-N., {Li}, M.-H., \& {Li}, X. 2018{\natexlab{a}}, \mnras, 480, 3117

\bibitem[{{Lin} {et~al.}(2018{\natexlab{b}}){Lin}, {Li}, \&
  {Li}}]{2018EPJC...78..356L}
{Lin}, H.-N., {Li}, J., \& {Li}, X. 2018{\natexlab{b}}, European Physical
  Journal C, 78, 356

\bibitem[{{Lv} \& {Xia}(2016)}]{2016PDU....13..139L}
{Lv}, M.-Z., \& {Xia}, J.-Q. 2016, Physics of the Dark Universe, 13, 139

\bibitem[{{Ma} \& {Corasaniti}(2018)}]{2018ApJ...861..124M}
{Ma}, C., \& {Corasaniti}, P.-S. 2018, \apj, 861, 124

\bibitem[{{Marriner} {et~al.}(2011){Marriner}, {Bernstein}, {Kessler},
  {Lampeitl}, {Miquel}, {Mosher}, {Nichol}, {Sako}, {Schneider}, \&
  {Smith}}]{2011ApJ...740...72M}
{Marriner}, J., {Bernstein}, J.~P., {Kessler}, R., {et~al.} 2011, \apj, 740, 72

\bibitem[{{Melia}(2018)}]{2018MNRAS.481.4855M}
{Melia}, F. 2018, \mnras, 481, 4855

\bibitem[{{Meng} {et~al.}(2012){Meng}, {Zhang}, {Zhan}, \&
  {Wang}}]{2012ApJ...745...98M}
{Meng}, X.-L., {Zhang}, T.-J., {Zhan}, H., \& {Wang}, X. 2012, \apj, 745, 98

\bibitem[{{Messenger} {et~al.}(2014){Messenger}, {Takami}, {Gossan},
  {Rezzolla}, \& {Sathyaprakash}}]{2014PhRvX...4d1004M}
{Messenger}, C., {Takami}, K., {Gossan}, S., {Rezzolla}, L., \&
  {Sathyaprakash}, B.~S. 2014, Physical Review X, 4, 041004

\bibitem[{{More} {et~al.}(2009){More}, {Bovy}, \& {Hogg}}]{2009ApJ...696.1727M}
{More}, S., {Bovy}, J., \& {Hogg}, D.~W. 2009, \apj, 696, 1727

\bibitem[{{Nair} {et~al.}(2011){Nair}, {Jhingan}, \&
  {Jain}}]{2011JCAP...05..023N}
{Nair}, R., {Jhingan}, S., \& {Jain}, D. 2011, \jcap, 5, 023

\bibitem[{{Nair} {et~al.}(2012){Nair}, {Jhingan}, \&
  {Jain}}]{2012JCAP...12..028N}
{Nair}, R., {Jhingan}, S., \& {Jain}, D. 2012, \jcap, 12, 028

\bibitem[{{Nissanke} {et~al.}(2010){Nissanke}, {Holz}, {Hughes}, {Dalal}, \&
  {Sievers}}]{2010ApJ...725..496N}
{Nissanke}, S., {Holz}, D.~E., {Hughes}, S.~A., {Dalal}, N., \& {Sievers},
  J.~L. 2010, \apj, 725, 496

\bibitem[{{Perlmutter} {et~al.}(1999){Perlmutter}, {Aldering}, {Goldhaber},
  {Knop}, {Nugent}, {Castro}, {Deustua}, {Fabbro}, {Goobar}, {Groom}, {Hook},
  {Kim}, {Kim}, {Lee}, {Nunes}, {Pain}, {Pennypacker}, {Quimby}, {Lidman},
  {Ellis}, {Irwin}, {McMahon}, {Ruiz-Lapuente}, {Walton}, {Schaefer}, {Boyle},
  {Filippenko}, {Matheson}, {Fruchter}, {Panagia}, {Newberg}, {Couch}, \&
  {Project}}]{1999ApJ...517..565P}
{Perlmutter}, S., {Aldering}, G., {Goldhaber}, G., {et~al.} 1999, \apj, 517,
  565

\bibitem[{{Punturo} {et~al.}(2010){Punturo}, {Abernathy}, {Acernese}, \&
  et~al.}]{2010CQGra..27s4002P}
{Punturo}, M., {Abernathy}, M., {Acernese}, F., \& et~al. 2010, Classical and
  Quantum Gravity, 27, 194002

\bibitem[{{Qi} {et~al.}(2019){Qi}, {Cao}, {Pan}, \& {Li}}]{2019arXiv190201702Q}
{Qi}, J.-Z., {Cao}, S., {Pan}, Y., \& {Li}, J. 2019, arXiv e-prints,
  arXiv:1902.01702

\bibitem[{{Rana} {et~al.}(2016){Rana}, {Jain}, {Mahajan}, \&
  {Mukherjee}}]{2016JCAP...07..026R}
{Rana}, A., {Jain}, D., {Mahajan}, S., \& {Mukherjee}, A. 2016, \jcap, 7, 026

\bibitem[{{Riess} {et~al.}(1998){Riess}, {Filippenko}, {Challis},
  {Clocchiatti}, {Diercks}, {Garnavich}, {Gilliland}, {Hogan}, {Jha},
  {Kirshner}, {Leibundgut}, {Phillips}, {Reiss}, {Schmidt}, {Schommer},
  {Smith}, {Spyromilio}, {Stubbs}, {Suntzeff}, \&
  {Tonry}}]{1998AJ....116.1009R}
{Riess}, A.~G., {Filippenko}, A.~V., {Challis}, P., {et~al.} 1998, \aj, 116,
  1009

\bibitem[{{Ruan} {et~al.}(2018){Ruan}, {Melia}, \&
  {Zhang}}]{2018ApJ...866...31R}
{Ruan}, C.-Z., {Melia}, F., \& {Zhang}, T.-J. 2018, \apj, 866, 31

\bibitem[{{Sathyaprakash} \& {Schutz}(2009)}]{2009LRR....12....2S}
{Sathyaprakash}, B.~S., \& {Schutz}, B.~F. 2009, Living Reviews in Relativity,
  12, 2

\bibitem[{{Sathyaprakash} {et~al.}(2010){Sathyaprakash}, {Schutz}, \& {Van Den
  Broeck}}]{2010CQGra..27u5006S}
{Sathyaprakash}, B.~S., {Schutz}, B.~F., \& {Van Den Broeck}, C. 2010,
  Classical and Quantum Gravity, 27, 215006

\bibitem[{{Schneider} {et~al.}(2001){Schneider}, {Ferrari}, {Matarrese}, \&
  {Portegies Zwart}}]{2001MNRAS.324..797S}
{Schneider}, R., {Ferrari}, V., {Matarrese}, S., \& {Portegies Zwart}, S.~F.
  2001, \mnras, 324, 797

\bibitem[{{Schutz}(1986)}]{1986Natur.323..310S}
{Schutz}, B.~F. 1986, \nat, 323, 310

\bibitem[{{Scolnic} {et~al.}(2018){Scolnic}, {Jones}, {Rest}, {Pan},
  {Chornock}, {Foley}, {Huber}, {Kessler}, {Narayan}, {Riess}, {Rodney},
  {Berger}, {Brout}, {Challis}, {Drout}, {Finkbeiner}, {Lunnan}, {Kirshner},
  {Sanders}, {Schlafly}, {Smartt}, {Stubbs}, {Tonry}, {Wood-Vasey}, {Foley},
  {Hand}, {Johnson}, {Burgett}, {Chambers}, {Draper}, {Hodapp}, {Kaiser},
  {Kudritzki}, {Magnier}, {Metcalfe}, {Bresolin}, {Gall}, {Kotak}, {McCrum}, \&
  {Smith}}]{2018ApJ...859..101S}
{Scolnic}, D.~M., {Jones}, D.~O., {Rest}, A., {et~al.} 2018, \apj, 859, 101

\bibitem[{{Uzan} {et~al.}(2004){Uzan}, {Aghanim}, \&
  {Mellier}}]{2004PhRvD..70h3533U}
{Uzan}, J.-P., {Aghanim}, N., \& {Mellier}, Y. 2004, \prd, 70, 083533

\bibitem[{{Wang} {et~al.}(2017){Wang}, {Wei}, {Li}, {Xia}, \&
  {Zhu}}]{2017ApJ...847...45W}
{Wang}, G.-J., {Wei}, J.-J., {Li}, Z.-X., {Xia}, J.-Q., \& {Zhu}, Z.-H. 2017,
  \apj, 847, 45

\bibitem[{{Wang} {et~al.}(2018){Wang}, {Zhang}, {Zhang}, \&
  {Zhang}}]{2018PhLB..782...87W}
{Wang}, L.-F., {Zhang}, X.-N., {Zhang}, J.-F., \& {Zhang}, X. 2018, Physics
  Letters B, 782, 87

\bibitem[{{Wei}(2018)}]{2018ApJ...868...29W}
{Wei}, J.-J. 2018, \apj, 868, 29

\bibitem[{{Wei} {et~al.}(2018){Wei}, {Wu}, \& {Gao}}]{2018ApJ...860L...7W}
{Wei}, J.-J., {Wu}, X.-F., \& {Gao}, H. 2018, \apjl, 860, L7

\bibitem[{{Yang} {et~al.}(2017){Yang}, {Holanda}, \&
  {Hu}}]{2017arXiv171010929Y}
{Yang}, T., {Holanda}, R.~F.~L., \& {Hu}, B. 2017, arXiv e-prints,
  arXiv:1710.10929

\bibitem[{{Yang} {et~al.}(2013){Yang}, {Yu}, {Zhang}, \&
  {Zhang}}]{2013ApJ...777L..24Y}
{Yang}, X., {Yu}, H.-R., {Zhang}, Z.-S., \& {Zhang}, T.-J. 2013, \apjl, 777,
  L24

\bibitem[{{Zhao} {et~al.}(2011){Zhao}, {van den Broeck}, {Baskaran}, \&
  {Li}}]{2011PhRvD..83b3005Z}
{Zhao}, W., {van den Broeck}, C., {Baskaran}, D., \& {Li}, T.~G.~F. 2011, \prd,
  83, 023005

\bibitem[{{Zhao} \& {Wen}(2018)}]{2018PhRvD..97f4031Z}
{Zhao}, W., \& {Wen}, L. 2018, \prd, 97, 064031

\bibitem[{{Zhao} {et~al.}(2018){Zhao}, {Wright}, \& {Li}}]{2018JCAP...10..052Z}
{Zhao}, W., {Wright}, B.~S., \& {Li}, B. 2018, \jcap, 10, 052

\end{thebibliography}

\end{document}